\documentclass[twocolumn,showpacs,preprintnumbers,amsmath,amssymb,superscriptaddress]{revtex4}

\usepackage{graphicx}
\usepackage{dcolumn}
\usepackage{bm}
\usepackage{color}

\begin{document}

\title{Force measurement in the presence of Brownian noise:\\
Equilibrium distribution method vs. Drift method}

\author{Thomas Brettschneider}
\affiliation{2. Physikalisches Institut, Universit\"{a}t Stuttgart, Pfaffenwaldring 57, 70550 Stuttgart, Germany}

\author{Giovanni Volpe}
\email{g.volpe@physik.uni-stuttgart.de}
\affiliation{2. Physikalisches Institut, Universit\"{a}t Stuttgart, Pfaffenwaldring 57, 70550 Stuttgart, Germany}
\affiliation{Max-Planck-Institut f\"{u}r Metallforschung, Heisenbergstra{\ss}e 3, 70569 Stuttgart, Germany}

\author{Laurent Helden}
\affiliation{2. Physikalisches Institut, Universit\"{a}t Stuttgart, Pfaffenwaldring 57, 70550 Stuttgart, Germany}

\author{Jan Wehr}
\affiliation{Department of Mathematics, University of Arizona, Tucson, AZ 85721-0089, USA}

\author{Clemens Bechinger}
\affiliation{2. Physikalisches Institut, Universit\"{a}t Stuttgart, Pfaffenwaldring 57, 70550 Stuttgart, Germany}
\affiliation{Max-Planck-Institut f\"{u}r Metallforschung, Heisenbergstra{\ss}e 3, 70569 Stuttgart, Germany}

\date{\today}

\begin{abstract}
The study of microsystems and the development of nanotechnologies require new techniques to measure piconewton and femtonewton forces at microscopic and nanoscopic scales. Amongst the challenges, there is the need to deal with the ineluctable thermal noise, which, in the typical experimental situation of a spatial diffusion gradient, causes a spurious drift. This leads to a correction term when forces are estimated from drift measurements [Phys. Rev. Lett. \textbf{104}, 170602 (2010)]. Here, we provide a systematic study of such effect comparing the forces acting on various Brownian particles derived from equilibrium distribution and drift measurements. We discuss the physical origin of the correction term, its dependence on wall distance, particle radius, and its relation to the convention used to solve the respective stochastic integrals. Such correction term becomes more significant for smaller particles and is predicted to be in the order of several piconewtons for particles the size of a biomolecule.
\end{abstract}


\maketitle

\section{Introduction}

The precise measurement of small forces plays a central role in science and technology. Apart from the instrumental challenge of measuring forces in the pico- and femtonewton range, the underlying concept of how forces are determined between macroscopic objects cannot simply be scaled down to the micro- and nanoscale. This is easily understood by considering, e.g., a micron-sized object suspended in a liquid environment. Due to its Brownian motion resulting from collisions with the solvent's molecules, inertial effects become largely negligible and the trajectory will look rather different compared to macroscopic objects \cite{Purcell1977,HappelBrenner}. Due to the irregularities of this type of motion, various possible mathematical descriptions for the trajectory of a microscopic object exist, e.g., the stochastic differential equations suggested by It\^o and Stratonovich. Under many conditions all these descriptions lead to the identical physical interpretation. This is no longer true when the diffusion coefficient of the suspended particle becomes position-dependent, as this typically occurs close to other particles or to a wall. In such cases, different mathematical descriptions of the Brownian motion are not identical and at most one correctly describes the physical reality. Although the need for such corrections - usually referred to as \emph{spurious drift} - has been realized several decades ago, e.g., for numerical simulations \cite{Ermak1978,Ryter1981,Sancho1982,Clark1987}, they are seldom applied when analyzing experimental data \cite{Lancon2001,Behrens2003}. While such corrections are often small, we have recently demonstrated that in the presence of a spatial dependence of the diffusion coefficient the negligence of the correct spurious drift term can lead not only to the wrong amplitude but even the wrong sign of the forces acting on a micron-sized object suspended in a liquid environment, a situation often encountered, e.g., in biophysical experiments \cite{Volpe2010PRL}.

\begin{figure}
\includegraphics*[width=3.25in]{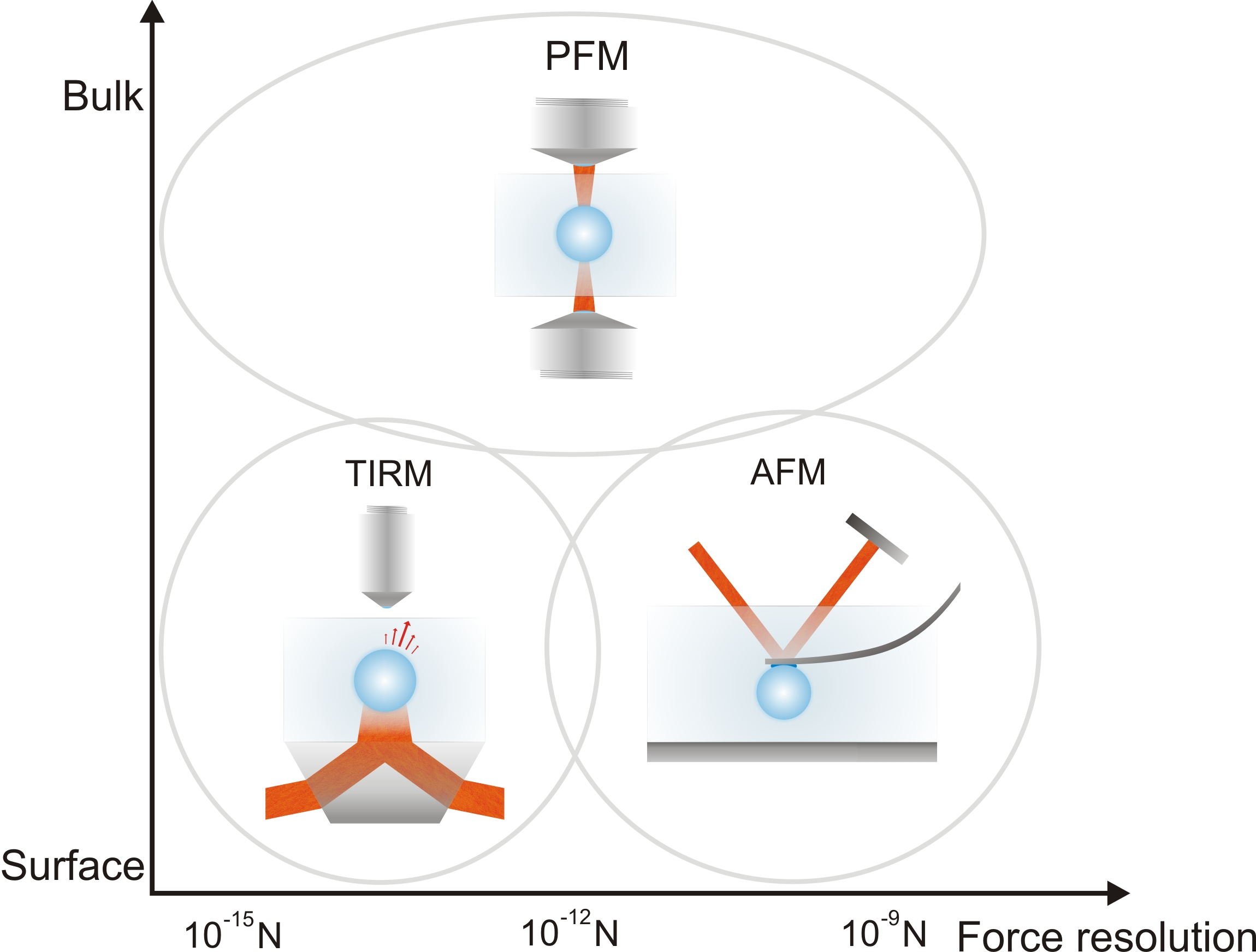}
\caption{Main techniques of measuring forces at the microscopic and nanoscopic scales, classified according to their force resolution and the working conditions -- surface/bulk -- for which they are best suited: atomic force microscopy (AFM)\cite{Binnig1986}, photonic force microscopy (PFM)\cite{Ghislain1993,BergSorensen2004,Volpe2007PRE1}, and total internal reflection microscopy (TIRM) \cite{Walz1997,Prieve1999,Volpe2009OE}. \label{fig:cdr_techniques}}
\end{figure}

In this article, we provide a systematic study of how the presence of a position-dependent diffusion coefficient alters the interpretation of the forces acting on a colloidal particle. We demonstrate that only when the correct spurious drift term is subtracted, the measured forces are in agreement with theoretical predictions. In particular, we show that the magnitude of the spurious forces increases when considering smaller particles as often employed in biophysical experiments: the spurious forces acting on a biomolecule with a diamter of about $10\,\mathrm{nm}$ can be on the order of several piconewtons. This value is comparable to the elastic forces that are commonly found in experiments probing the biomechanics of single molecules \cite{Mehta1999,Bustamante2003}.


In Section II we discuss the physical origin of the mathematical ambiguity in the description of Brownian motion in a diffusion gradient and how this ambiguity may result in different formulae for the estimation of a force from experimental data. We also show that there are essentially two ways of measuring forces  at the nanoscale. The first is based on the \emph{equilibrium distribution} of a particle (or particles) subjected to an \emph{a priori} unknown force. The second takes advantage of the fact that an applied force results in a \emph{drift} of the particle.
In Section III we provide some detailed information on the TIRM technique and the materials we employed. In Section IV, we report the results of the force measurements on particles of various sizes and materials suspended in water close to a wall under the action of electrostatic forces and effective gravity, quantifying in each case the value of the \emph{spurious forces}. In Appendix A, finally, we clarify how to describe correctly such phenomena using both stochastic differential equations and Fokker-Planck equations.

\section{Theory: Force measurements in a overdamped system}

The forces acting on a microscopic object immersed in a fluid medium can be assessed either by studying the underlying potential or by studying their effect on the object's trajectory (for an overview of force measurement techniques and their force resolution, see Fig.~\ref{fig:cdr_techniques}). The first approach -- to which we shall refer as \emph{equilibrium distribution method} -- requires sampling of the equilibrium distribution. Accordingly, it can be only applied under conditions where the investigated system is at or close to thermodynamic equilibrium with a heat bath.
The second method -- to which we shall refer as \emph{drift method} -- does not require the object to be at or even close to thermal equilibrium. Therefore, it can be applied also to systems which are
intrinsically out-of-equilibrium, e.g., molecular machines, transport through pores, DNA stretching \cite{Mehta1999,Bustamante2003}. The latter method, however, requires to detect the object trajectory with high sampling rates, which can be technologically more challenging, in particular when combined with a high spatial resolution.

\subsection{Equilibrium distribution method}

\begin{figure}
\includegraphics*[width=3.25in]{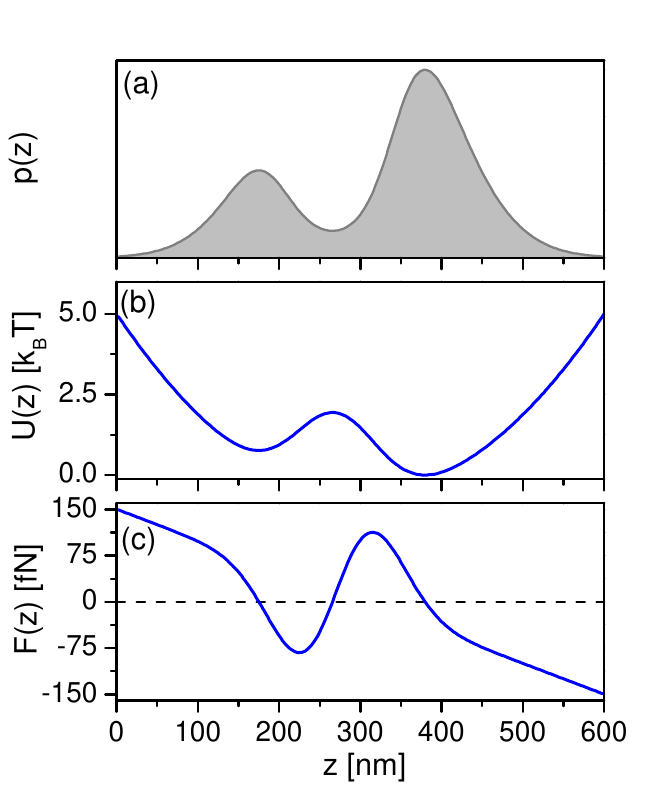}
\caption{Schematic procedure of measuring a force $F(z)$ by the equilibrium distribution method. From the measured equilibrium probability distribution $p(z)$ (a) of a particle one obtains the potential energy distribution $U(z)=-\ln p(z)$ (b), which then gives the force $F(z)=-\frac{\partial}{\partial z} U(z)$ (Eq.~\ref{eq:force_equilibrium}) (c). \label{fig:potential_method}}
\end{figure}

A microscopic object in contact with a thermal heat bath at constant temperature does not come to rest, but keeps on jiggling around due to the presence of thermal agitation. When the particle is subjected to an external potential well $U(z)$, this leads to a Boltzmann distribution $p(z) = \exp\left(-\frac{U(z)}{k_B T}\right)$, where $k_B$ is the Boltzmann constant and $T$ is the temperature of the heat bath.

Accordingly, it is possible to sample the steady-state position probability distribution $p(z)$ by measuring a large number of uncorrelated object positions (Fig.~\ref{fig:potential_method}(a)).
The equilibrium potential can be derived by $U(z) = -k_B T \ln(p(z))$ (Fig.~\ref{fig:potential_method}(b)) and the force (Fig.~\ref{fig:potential_method}(c)) by
\begin{equation}\label{eq:force_equilibrium}
F(z) = -\frac{dU(z)}{dz} = \frac{k_B T}{p(z)} \frac{dp(z)}{dz}.
\end{equation}

Due to the exponential dependence of the probability distribution on the potential depth, in typical experiments only potential minima within typically less than $\approx$ 5 $k_B T$ are explored by the particle. While Eq.~(\ref{eq:force_equilibrium}) can be only applied to equilibrium conditions, it should be mentioned that that it has recently been demonstrated that for the specific case of nonequilibrium steady states (NESS) a similar relation, i.e. a generalized Boltzmann distribution, can be derived for the stationary particle equilibrium distribution and the conservative part of the potential \cite{Blickle2007PRL,Blickle2007PRE}.

\subsection{Drift method}

\begin{figure}
\includegraphics*[width=3.25in]{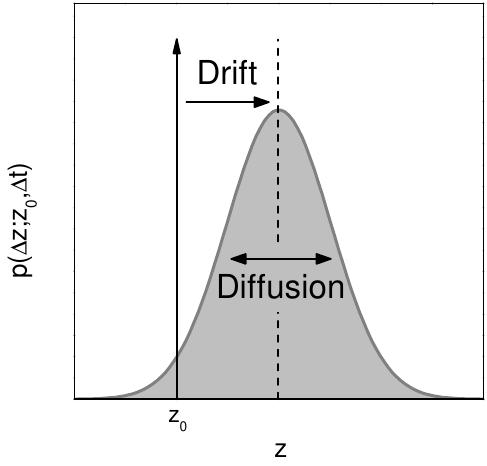}
\caption{Schematic view of the propagator $p(\Delta z;z_0,\Delta t)$, which gives the distribution of the particle position increments $\Delta z$ from its initial position $z_0$ after a time step $\Delta t$. The distribution is Gaussian for sufficiently small $\Delta t$ with standard deviation $\sqrt{2D(z_0)\Delta  t}$ (Eq.~(\ref{eq:diffusion})). The force can be estimated from the measured drift according to Eq.~(\ref{eq:force_nonequilibrium_rand}).
\label{fig:viscosity_drag_method}}
\end{figure}

Since for a microscopic body suspended in a liquid medium viscous forces prevail by several orders of magnitude over inertial effects, a constant force $F$ applied to a microscopic particle results in a constant drift velocity $v = F/\gamma$, where $\gamma$ is the object's friction coefficient.
Since $v = \Delta z /\Delta t$ can be retrieved from the measured particle displacement $\Delta z$ within time $\Delta t$, the force can be measured accordingly  as
\begin{equation}\label{eq:force_nonequilibrium_det}
F = \gamma \frac{\Delta z}{\Delta t}.
\end{equation}
For large forces this obviously leads to an univocal result. However, when the \emph{drift} force amplitude is comparable to the effect of the thermal noise, the measured particle displacement $\Delta z$ and, thus, the drift force vary between identical experiments, leading to a statistical \emph{distribution} of the measured values (Fig.~\ref{fig:viscosity_drag_method})
\begin{equation}\label{eq:principle_force}
F(z) = \gamma \left\langle\frac{\Delta z_j(z)}{\Delta t}\right\rangle.
\end{equation}
where $\Delta z_j(z)$ denotes the $j$-th experimental value of the particle's displacement after time $\Delta t$.

Although Eq.~(\ref{eq:principle_force}) is key to measure forces under nonequilibrium conditions, it is only valid in situations where the diffusion coefficient $D=kT/\gamma$ of the object
to which the force is applied is constant.
When $D$ becomes position-dependent, i.e.
\begin{equation}\label{eq:diffusion}
D(z) = \left\langle \frac{\left(\Delta z_j(z) - \langle \Delta z_j(z) \rangle\right)^2}{2\Delta t} \right\rangle,
\end{equation}
Eq.~(\ref{eq:principle_force}) must be corrected by an additional term \cite{Oksendal2007,Volpe2010PRL}
\begin{equation}\label{eq:force_nonequilibrium_rand}
F(z) = \gamma(z) \left\langle\frac{\Delta z_j(z)}{\Delta t}\right\rangle - \alpha \gamma(z) \frac{dD(z)}{dz}.
\end{equation}
to which we shall refer as \emph{spurious force} and which may depend on the specific choice of $\alpha \in [0,1]$.

In the case of systems that are strongly coupled to a heat bath, thermodynamic consistency requires that $\alpha = 1$ \cite{Lau2007}. In particular, this is true for a Brownian particle, which is the system we have experimentally investigated. More generally, other values of $\alpha$ might be possible when describing other stochastic processes \cite{Kupferman2004}.

\subsubsection{The physical origin of the spurious force}

\begin{figure}
\includegraphics*[width=3.25in]{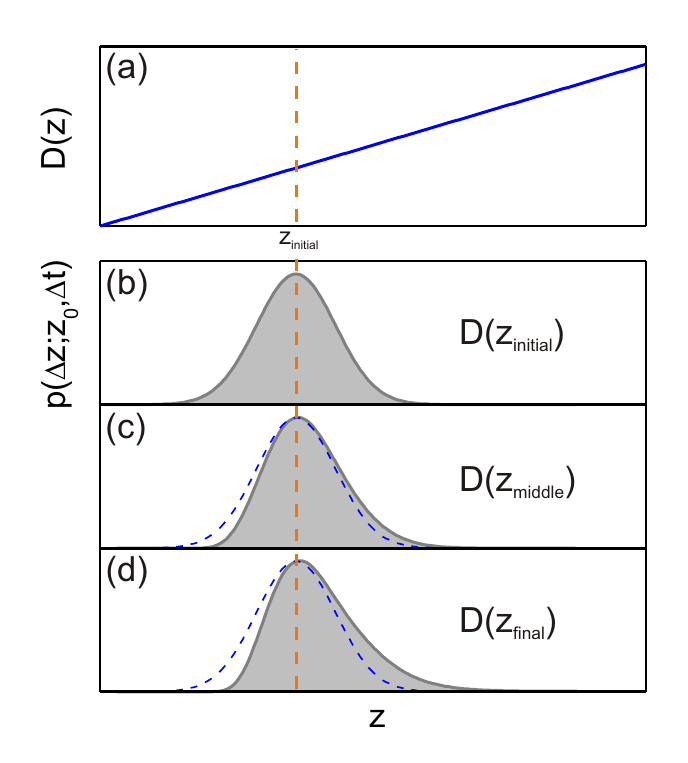}
\caption{In the presence of a diffusion gradient (a), the propagator $p(\Delta z;z_0,\Delta t)$ is different depending on where $D$ is evaluated, i.e. at $z_{\text{initial}}$ (b), at $z_{\text{middle}}$ (c), or at $z_{\text{final}}$ (d). In the latter two cases this results in a shift of the position distribution mean -- the spurious drift.\label{fig:distorted_distributions}}
\end{figure}

A qualitative physical understanding of the correction term in Eq.~(\ref{eq:force_nonequilibrium_rand}) can be gained considering the effect of a diffusion gradient (Fig.~\ref{fig:distorted_distributions}(a)) on a Brownian particle initially localized at position $z_{\mathrm{initial}}$ (orange dashed line in Fig.~\ref{fig:distorted_distributions}) at time $t_0$. In the simplest picture, the particle diffusion results in a dichotomic movement either to the left or to the right with the same probability; therefore after time $\Delta t$ the particle is displaced by $z_{\mathrm{initial}} \pm \sqrt{2 D} \Delta t$. In a more realistic picture, the particle final position has a continuous probability distribution. In both cases, assuming $D$ constant, the final particle position distribution $p(\Delta z; z_0, \Delta t)$ is symmetric, such as the histogram of Fig.~\ref{fig:distorted_distributions}(b).

In the presence of a diffusion gradient (Fig.~\ref{fig:distorted_distributions}(a)) the value of $D$ is obviously different at the initial and final position and, therefore, the evaluation of the displacement is not univocal. Assuming $D = D(z_\mathrm{intial})$ -- to which we shall refer as \emph{It\^o convention} -- $p(\Delta z; z_0, \Delta t)$ is symmetric, as in the constant diffusion case (Fig.~\ref{fig:distorted_distributions}(b)). However, it could be argued that $D$ should be averaged over the particle displacement; assuming thus $D = D(z_\mathrm{middle})$ -- \emph{Stratonovich convention} -- $p(\Delta z; z_0, \Delta t)$ becomes asymmetric (Fig.~\ref{fig:distorted_distributions}(c)), because the particle displaces further when moving towards increasing diffusion. Finally, assuming $D = D(z_\mathrm{final})$ -- \emph{isothermal}, \emph{anti-It\^{o}} or \emph{backwards-It\^o convention} -- $p(\Delta z; z_0, \Delta t)$ becomes even more asymmetric (Fig.~\ref{fig:distorted_distributions}(d)). The \emph{spurious drift}, and the related \emph{spurious force}, account for such asymmetry.

In more general terms, we might assume $D = (1-\alpha) D(z_{\mathrm{initial}}) + \alpha D(z_{\mathrm{final}})$ and $\alpha \in [0,1]$, with $\alpha = 0 $ corresponding to the It\^{o} convention, $\alpha = 0.5$ to the Stratonovich convention, and $\alpha = 1$ to the isothermal convention. The latter one, in particular, turns out to be the only thermodynamically consistent one for a system coupled to a heat bath \cite{Lau2007}.

\section{Methods and materials}

\subsection{Total internal reflection microscopy (TIRM)}\label{sec:tirm}

\begin{figure}
\includegraphics*[width=3.25in]{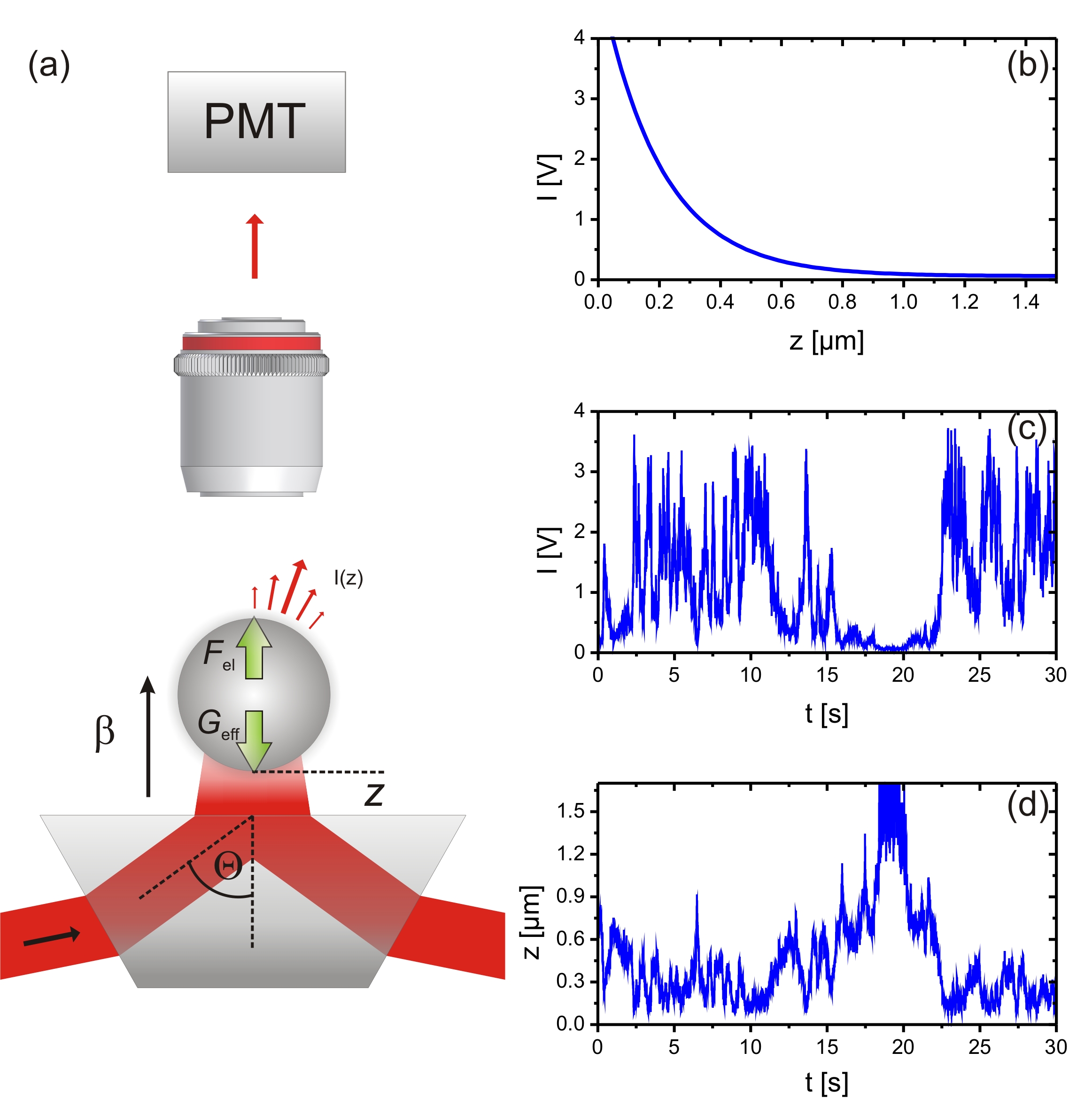}
\caption{Total internal reflection microscopy (TIRM). (a) Sketch of a typical TIRM geometry with a single colloidal particle in front of a planar transparent wall. The particle is illuminated by the evanescent field created by total internal reflection of a laser beam at the glass-fluid interface while performing Brownian motion. The scattered intensity $I$ is measured using a photomultiplier (PMT). The forces acting on the particle are due to gravity $G_{\mathrm{eff}}$ and electrostatic interactions $F_{\mathrm{el}}$ and are indicated by green arrows. When the relationship between the scattered intensity and distance $I(z)$ is known (b) the intensity time-series $I(t)$ (c) can be converted into the trajectory $z(t)$ (d).
\label{fig:tirm_principle}}
\end{figure}

The trajectories of colloidal particles close to a wall are measured with TIRM \cite{Walz1997,Prieve1999,Volpe2009OE} and a scheme of a typical setup is presented in Fig.~\ref{fig:tirm_principle}(a). A p-polarized laser beam ($\lambda = 658~\mathrm{nm}$) is totally internally reflected at a glass-liquid interface generating an evanescent field decaying into the liquid. A spherical colloidal particle in the vicinity of the interface scatters the evanescent light. The scattering intensity shows a marked dependence on the particle position $I(z) = I_0 \exp(-\beta z )$ where $\beta$ is the inverse evanescent decay length and $I_0 = I(z=0)$ (Fig.~\ref{fig:tirm_principle}(b)). The scattering intensity timeseries $I(t)$ is collected by an objective and recorded by a photomultiplier (Fig.~\ref{fig:tirm_principle}(c)). Inverting the position-intensity relation, we finally obtain the particles trajectory $z(t)$ with a spatial resolution of a few nanometers (Fig.~\ref{fig:tirm_principle}(d)). Since our system is in thermal equilibrium, the measured particle trajectory allows us to calculate the forces via both the equilibrium distribution and the drift method, as described above.

\begin{figure}
\includegraphics*[width=3.25in]{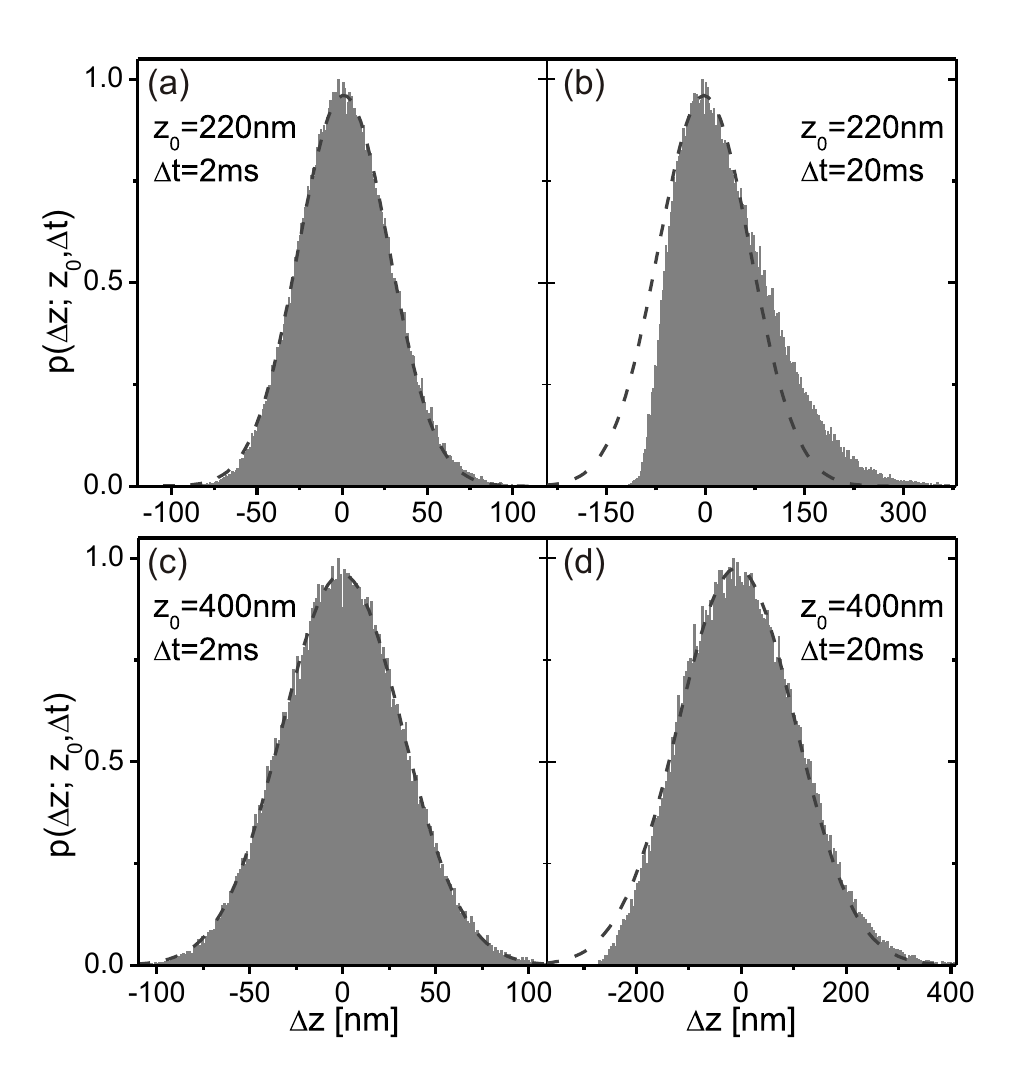}
\caption{Influence of $\Delta t$. At the distance $z_0=220\,\mathrm{nm}$ $\Delta t = 2\,\mathrm{ms}$ leads to a Gaussian distribution of the position increments (a), but for $\Delta t=20\,\mathrm{ms}$ a clear deviation from a Gaussian distribution appears. At $z = 400\,\mathrm{nm}$, where the spatial variation of the drift is much smaller, the distribution is again Gaussian for $\Delta t = 2\,\mathrm{ms}$ (c) and remains almost Gaussian also for $\Delta t = 20\,\mathrm{ms}$.\label{fig:skewness}}
\end{figure}

Due to the fractal nature of the Brownian motion, the value of $\left\langle\frac{\Delta z_j(z)}{\Delta t}\right\rangle$ which has to be calculated for the drift force (see Eq.~(\ref{eq:force_nonequilibrium_rand})) strongly depends on the chosen time interval $\Delta t$. On the one hand $\Delta t$ should be as short as possible, on the other hand it cannot be made arbitrarily small due to the finite experimental acquisition frequency. A reasonable tradeoff is to choose $\Delta t$ so small that the spatial variation of the drift force during such a time step is negligible. In order to meet these conditions in our experiment, we have stepwise reduced $\Delta t$ until the probability distribution of $\Delta z$ became Gaussian, as expected when the above condition holds. Because under our conditions (see next paragraph) the spatial gradient of the force acting on the particle is not constant, the above condition on $\Delta t$ is expected to vary with the particle-wall distance. This is shown in Fig.~\ref{fig:skewness} where we plotted the probability distribution of $\Delta z$ for $\Delta t = 2\,\mathrm{ms}$ and $\Delta t = 20\,\mathrm{ms}$ obtained at two particle-wall distances $z = 220\,\mathrm{nm}$ and $z = 400\,\mathrm{nm}$ which correspond to the distances with the largest and smallest spatial variation of the force acting on the particle. As can be seen, for $\Delta t = 2\,\mathrm{ms}$, we always obtain a Gaussian distribution while for $\Delta t = 20\,\mathrm{ms}$ deviations from a Gaussian fit (dashed line) are observed. In all measurements presented in the following we have chosen $\Delta t = 2\,\mathrm{ms}$.

\subsection{Gravitational and electrostatic force}

In order to compare the measured forces with theory, it is important to have full knowledge about the interaction mechanisms of a colloidal particle close to a wall.
For an electrically charged dielectric colloidal sphere suspended in a solvent, the interaction forces have been demonstrated to be described by \cite{Walz1997,Prieve1999,Volpe2009OE}
\begin{equation}\label{eq:physical_forces}
F(z) = Be^{-\kappa z} - G_{\mathrm{eff}}.
\end{equation}
The first term is due to double layer forces with $\kappa^{-1}$ the Debye length and $B$ a prefactor depending on the surface charge densities of the particle and the wall (see Tab.~\ref{tab:sample_table}). The second term describes the effective gravitational contributions $G_{\mathrm{eff}} = \frac{4}{3} \pi R^3 (\rho_p - \rho_s) g$ with $\rho_p$ and $\rho_s$ the particle and solvent density and $g$ the gravitational acceleration constant. Under our conditions the additional contribution of van der Waals forces can be neglected since they become relevant only at much shorter particle-wall distances \cite{Bevan1999,Grunberg2001}.

\subsection{Diffusion in bulk and diffusion gradient in front of a wall}

\begin{figure}
\includegraphics*[width=3.25in]{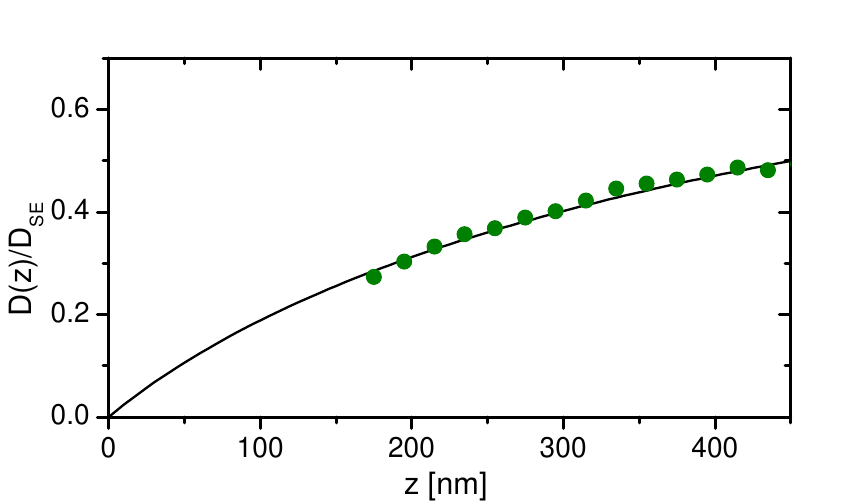}
\caption{Comparison of measured (bullets) and calculated (line) normalized vertical diffusion coefficient $D_{\bot}/D_{\mathrm{SE}}$ for an $R = 400\,\mathrm{nm}$ particle as a function of the particle-wall separation $z$.
\label{fig:D}}
\end{figure}

The Stokes-Einstein diffusion coefficient of a spherical colloidal particle immersed in a solvent is $D_{\mathrm{SE}} = k_BT/6\pi\eta R$, where $\eta$ is the shear viscosity of the liquid.
Close to a wall the bulk diffusion coefficient decreases due to hydrodynamic interactions. From the solution of the creeping flow equations for a spherical particle moving near a wall assuming nonslip boundary conditions, the vertical component of the diffusion
coefficient is \cite{Brenner1961}
\begin{equation}\label{eq:brenner}
D_{\bot}(z) = \frac{D_{\mathrm{SE}}}{l(z)} \mbox{,}
\end{equation}
where $l(z) = \frac{4}{3}\sinh{\left( a(z) \right)}
\sum_{n=1}^{\infty} \frac{n(n+1)}{(2n-1)(2n+3)}$ $\left[ \frac{
2\sinh{\left( (2n+1) a(z) \right)} + (2n+1)\sinh{\left(
2 a(z) \right)} }{ 4\sinh^2{\left( (n+0.5) a(z) \right)} -
(2n+1)^2\sinh^2{\left( a(z) \right)}} -1 \right]$ and
$a(z) = \cosh^{-1}\left( 1+\frac{z}{R}\right)$. $D_{\bot}(z)$ is zero at the wall and monotonically increases with $z$ approaching the bulk value at a distance of several
particle radii away from the wall. When calculating $D_{\bot}(z)$ from the particle trajectories of our TIRM measurements according to Eq.~(\ref{eq:diffusion}), indeed we find good agreement with the theoretical prediction (Fig.~\ref{fig:D}).

\subsection{Sample preparation and parameters}

\begin{table*}
\caption{\label{tab:sample_table}Sample parameters.}
\begin{ruledtabular}
\begin{tabular}{ccccccc}
$R~\mathrm{[nm]}$ &
Material &
$\rho_s~\mathrm{[g/cm^3]}$ &
$\kappa^{-1}~\mathrm{[nm]}$ &
$B~\mathrm{[pN]}$ &
NaCl $\mathrm{[\mu m]}$ \\
\hline
\hline
400 &
titanium-oxide &
2.54 &
25 &
68 &
150 \\
655 &
melamin &
1.51 &
18 &
770 &
300 \\
1180 &
polystyrene &
1.05 &
18 &
1080 &
300\\
\end{tabular}
\end{ruledtabular}
\end{table*}

As sample cell we used a $2~\mathrm{mm}$ thick cuvette comprising two optical flats separated by a spacer of silicon rubber.
The cell was filled with clean deionized water containing a very small number of colloidal particles and $150-300~\mathrm{\mu mM}$ NaCl salt to adjust the Debye screening length. In order to vary the spatial gradients of the diffusion coefficient we used particles with different radii $R$ and densities since they sample different ranges of $z/R$: $R=400\,\mathrm{nm}$ (titanium oxide), $R=655\,\mathrm{nm}$ (melamin), and $R=1180\,\mathrm{nm}$ (polystyrene). For further details see Tab.~\ref{tab:sample_table}.

\section{Results and discussion}

\begin{figure}
\includegraphics*[width=3.25in]{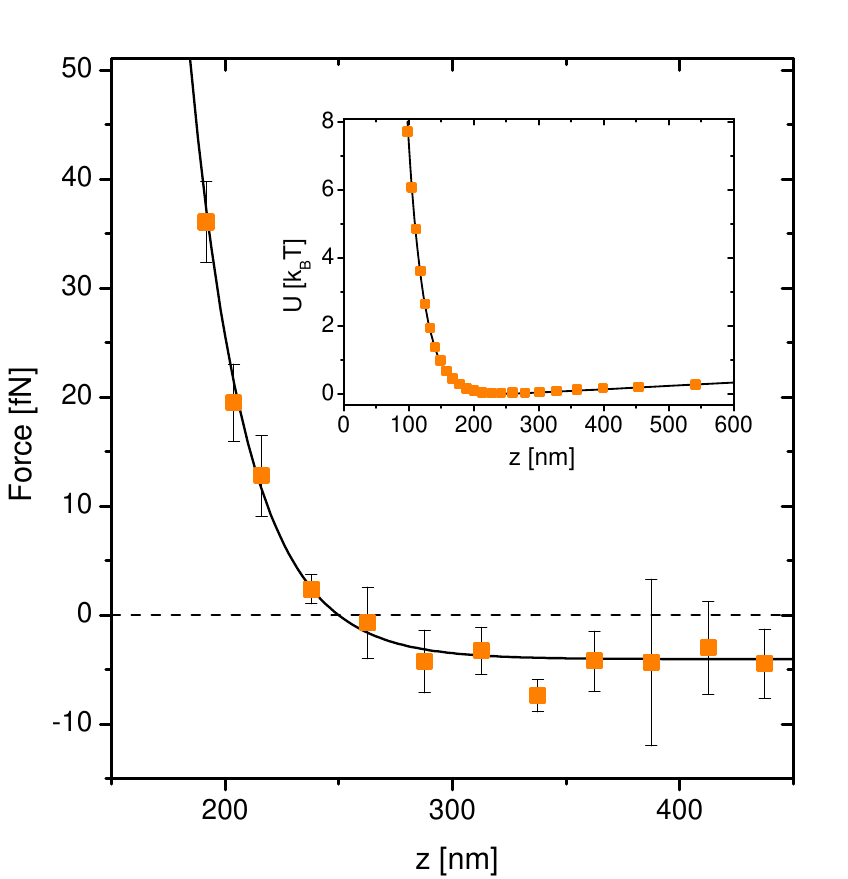}
\caption{Force derived by the equilibrium distribution method (squares) for the $400\,\mathrm{nm}$ radius particle (see Tab.~\ref{tab:sample_table}) and theoretical expectation (line) according to Eq.~(\ref{eq:physical_forces}). The inset shows the corresponding potential $U(z)$.\label{fig:result_tio2_pot}}
\end{figure}

In Fig.~\ref{fig:result_tio2_pot} we show as closed symbols the force (and as inset the corresponding potential U(z)) acting on a $R = 400\,\mathrm{nm}$ titanium oxide particle as obtained by the equilibrium distribution method (Eq.~(\ref{eq:force_equilibrium})). Since the forces as determined from this method are unambiguous, they will be considered as the true forces acting on the particle. This is also supported by the fact that the experimental data are in quantitative agreement with Eq.~(\ref{eq:physical_forces}) (solid line), where $G_{\mathrm{eff}}$ and $\kappa^{-1}$ are taken from the experimentally known parameters (Tab.~\ref{tab:sample_table}) while the prefactor $B$ has been treated as a fit parameter. The value $B = 68 \, \mathrm{pN}$ is in agreement with other TIRM measurements under similar conditions \cite{Prieve1999}.

\begin{figure}
\includegraphics*[width=3.25in]{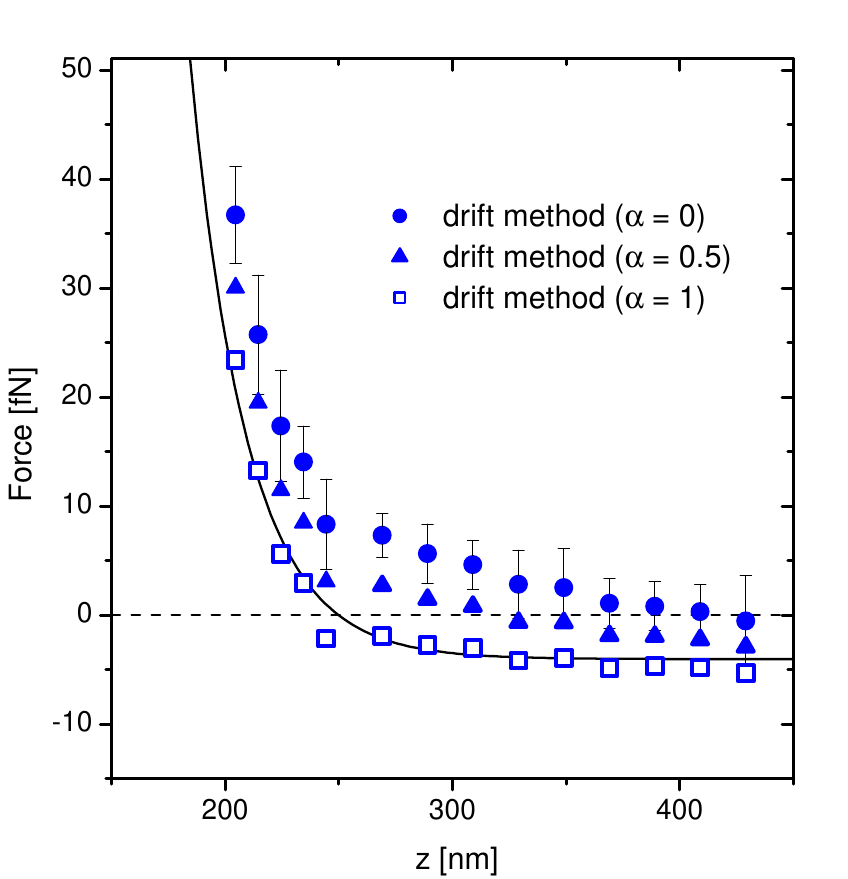}
\caption{Force derived by the drift method for the very same particle as in Fig. \ref{fig:result_tio2_pot}. The force measured according to Eq.~(\ref{eq:force_nonequilibrium_rand}) with $\alpha = 1$ (bullets) coincides with the theory (line), while with $\alpha = 0$ (circles) and $\alpha = 0.5$ (triangles) there is a clear disagreement. \label{fig:result_tio2_drift}}
\end{figure}

The symbols in Fig.~\ref{fig:result_tio2_drift} correspond to the force-distance relation as obtained from the drift method using Eq.~(\ref{eq:force_nonequilibrium_rand}) for $\alpha = 0$ (bullets), $\alpha = 0.5$ (triangles) and $\alpha = 1$ (open squares). Since the gradient of $D$ vanishes far away from the surface, the force dependence on $\alpha$ is most pronounced close to the wall but weakens at larger $z$ where the curves will eventually merge (only beyond the maximum distance sampled by the particle).
The forces determined with $\alpha = 1$, i.e. the isothermal convention, show good agreement with the Eq.~(\ref{eq:physical_forces}) (solid line, the same as in Fig.~\ref{fig:result_tio2_pot}). We want to emphasize that all other choices of $\alpha$, in particular negligence of the noise-induced correction ($\alpha = 0$), lead to significant differences. Not only the magnitude but also the sign of the forces obtained with $\alpha = 0, 0.5$ disagree with the true forces as obtained in Fig.~\ref{fig:result_tio2_pot}.

\begin{figure}
\includegraphics*[width=3.25in]{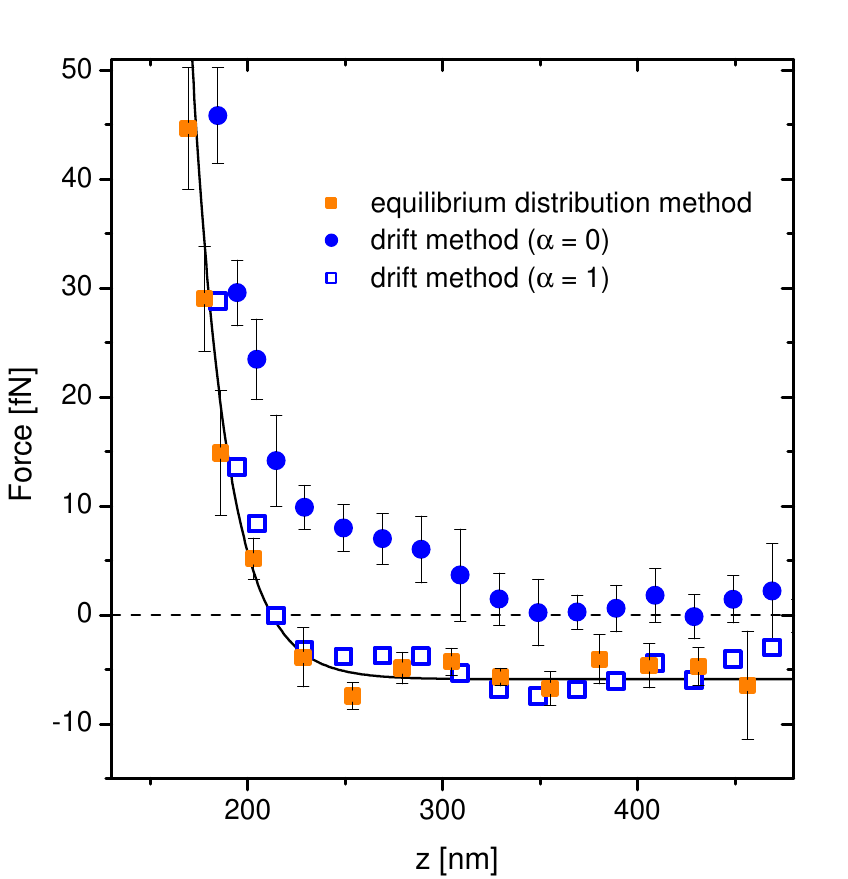}
\caption{Comparison of forces obtained for the $655\,\mathrm{nm}$ radius particle (see Tab.~\ref{tab:sample_table}). For the equilibrium distribution method (closed squares) and drift method with correction term (open squares) the results are in agreement with the theoretical expectation (line) according to Eq.~(\ref{eq:physical_forces}). Neglecting the correction term leads to a clear deviation (bullets).\label{fig:result_melamin}}
\end{figure}

\begin{figure}
\includegraphics*[width=3.25in]{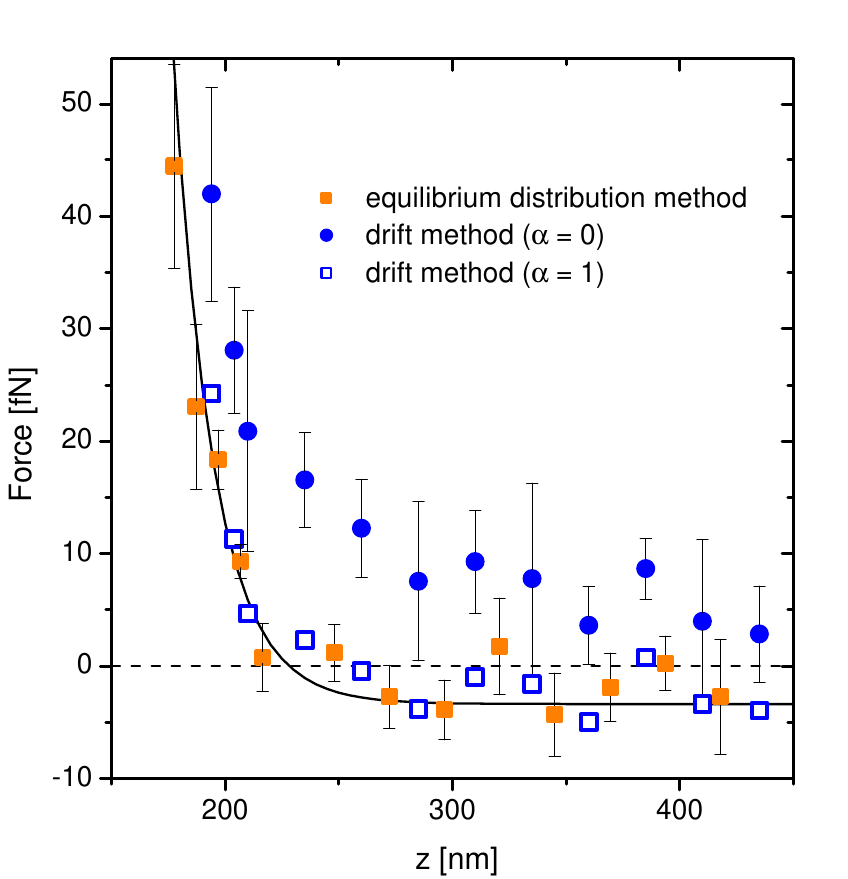}
\caption{Comparison of forces obtained for the $1180\,\mathrm{nm}$ radius particle (see Tab.~\ref{tab:sample_table}). For the equilibrium distribution method (closed squares) and drift method with correction term (open squares) the results are in agreement with the theoretical expectation (line) according to Eq.~(\ref{eq:physical_forces}). Neglecting the correction term leads to a clear deviation (bullets).\label{fig:result_ps}}
\end{figure}

\begin{figure}
\includegraphics*[width=3.25in]{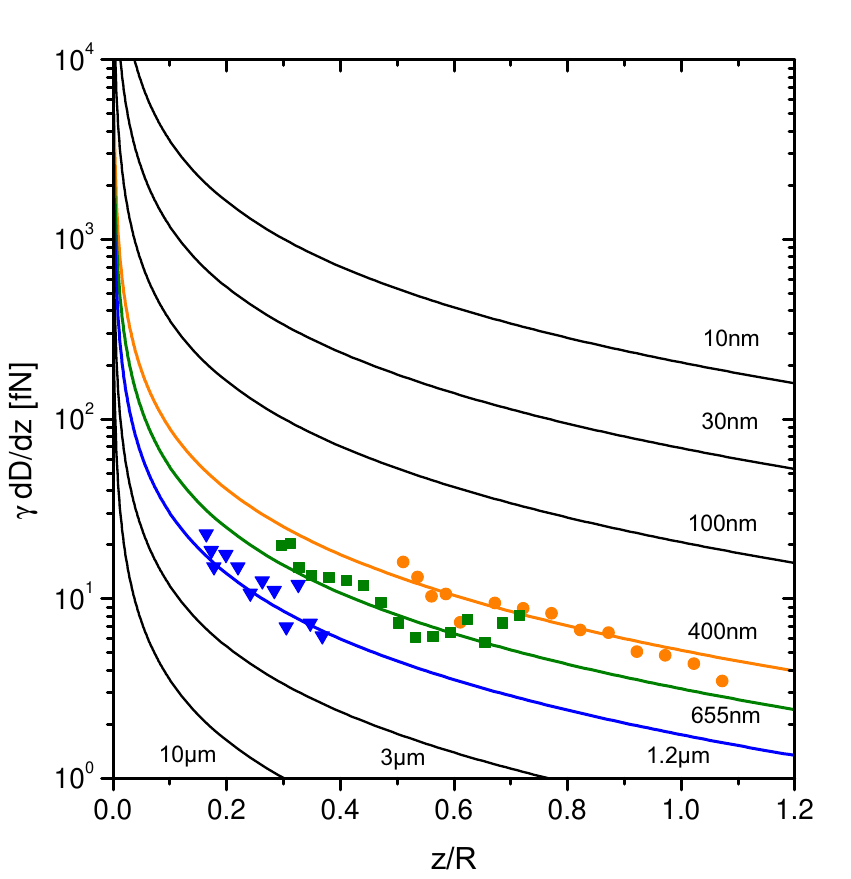}
\caption{Distance dependence of the theoretically calculated spurious force $\alpha \gamma\, \frac{d}{dz}D$ with $\alpha = 1$ for various particle radii $R$ (solid lines). Experimental data is shown for $R = 400\,\mathrm{nm}$ (circles, cf.~Fig.~\ref{fig:result_tio2_drift}), $R = 655\,\mathrm{nm}$ (squares, cf.~Fig.~\ref{fig:result_melamin}), and $R = 1180\,\mathrm{nm}$ (triangles, cf.~Fig.~\ref{fig:result_ps}).\label{fig:correction_term_force}}
\end{figure}

\begin{figure}
\includegraphics*[width=3.25in]{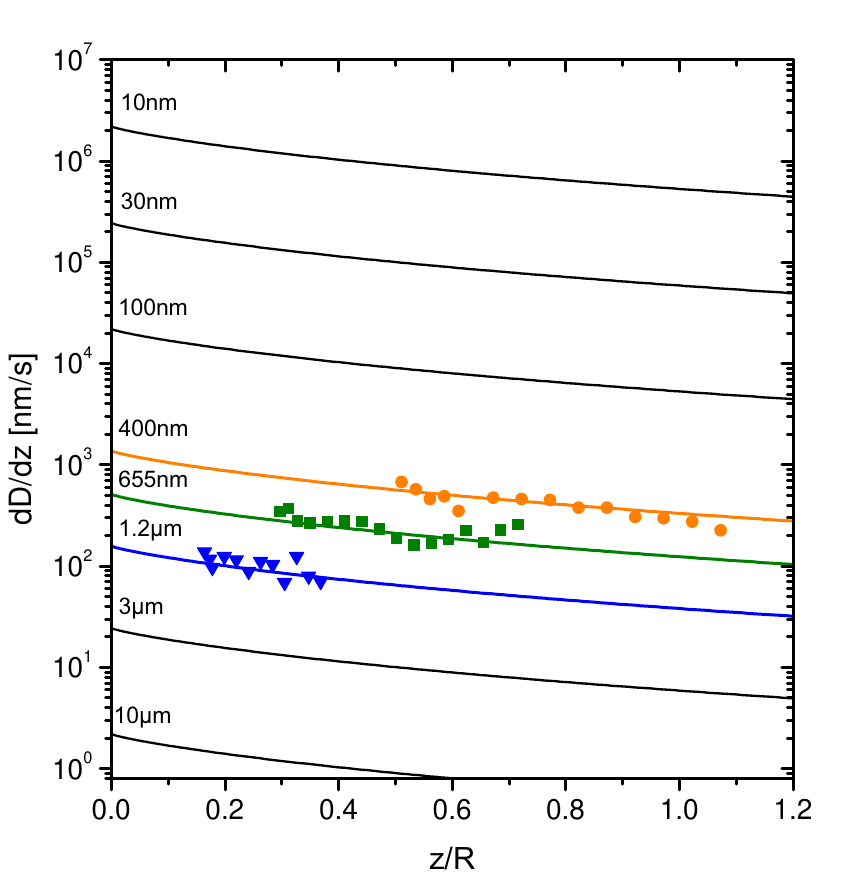}
\caption{Distance dependence of the spurious drift $\frac{d}{dz}D$ for various particle radii $R$ (solid lines). Experimental data is shown for $R = 400\,\mathrm{nm}$ (circles, cf.~Fig.~\ref{fig:result_tio2_drift}), $R = 655\,\mathrm{nm}$ (squares, cf.~Fig.~\ref{fig:result_melamin}), and $R = 1180\,\mathrm{nm}$ (triangles, cf.~Fig.~\ref{fig:result_ps}).\label{fig:correction_term_drift}}
\end{figure}

Similar measurements were also performed with other particles which are capable to sample even smaller particle-wall distance normalized by the particle radius, i.e. $z/R$ (cf. Figs.~\ref{fig:correction_term_force} and \ref{fig:correction_term_drift}), where the gradient of the diffusion coefficient becomes larger. Fig.~\ref{fig:result_melamin} shows the results for the $R=655\,\mathrm{nm}$ particle and Fig.~\ref{fig:result_ps} for the $R = 1080\,\mathrm{nm}$ particle. In both cases, the equilibrium distribution measurement (orange squares) agrees with Eq.~(\ref{eq:physical_forces}) with $B$ as the only fit parameter and with $\alpha = 1$ (blue open squares). As before, all other choices of $\alpha$ show no agreement, as exemplarily plotted for $\alpha = 0$ (blue bullets).

In Fig.~\ref{fig:correction_term_force}, the experimentally determined spurious force obtained by the difference between the forces derived with the uncorrected drift method (Eq.~(\ref{eq:principle_force}) or Eq.~(\ref{eq:force_nonequilibrium_rand}) with $\alpha =0$) and the equilibrium potential method (Eq.~(\ref{eq:force_equilibrium})) is plotted for different particle sizes as a function of $z/R$. The data are in good agreement with the solid lines representing the theoretical predictions for the spurious force $\alpha \gamma \frac{d}{dz}D(z)$ with $\alpha = 1$.
The spurious force depends only on the particle radius, but is independent of $\rho_s$. It increases for particles nearer to the wall and for smaller particles due to the larger $D$, reaching values in the order of several piconewtons for particles with $R=10\,\mathrm{nm}$, the size of a macromolecule.

Dividing the spurious forces (Fig.~\ref{fig:correction_term_force}) by $\gamma$, it is possible to derive the spurious drift $\alpha \frac{d}{dz}D(z)$, which is plotted in Fig.~\ref{fig:correction_term_drift} for $\alpha = 1$. The lines corresponding to the theoretical predictions and the symbols representing the experimental data for the particles with $R=400\,\mathrm{nm}$ (orange bullets), $R=655\,\mathrm{nm}$ (green squares) and $R=1080\,\mathrm{nm}$ (blue triangles) show good agreement. Due to the increasing gradient of the diffusion coefficient when approaching a surface, the spurious drift increases for shorter particle-wall distances and for smaller particles. Differently form the spurious force, which diverges for small $z$, the spurious drift reaches a maximum at $z=0$.

\section{Conclusions}

The experiments presented in this article clearly demonstrate that the presence of noise-induced drift has to be considered in force measurements based on the drift; neglecting it, this can lead to artifacts, which may even suggest the wrong sign of the force. Such spurious forces become more significant for smaller systems and may reach the piconewton range for objects the size of a biomolecule, i.e. about 10 nanometers.
We stress, furthermore, that a constant diffusion can be assumed only for a particle far from any boundary. Such boundaries are naturally introduced by surfaces or by other particles in suspension, a situation that is typically met in experiments.

\appendix

\section{Mathematical description of the spurious drift: stochastic differential equations and Fokker-Planck equation}

The motion of a Brownian particle can be described in several ways. In this Appendix we clarify the relations between the
various approaches, putting them in a historical context and demonstrating how the spurious drift emerges. We refer to Ref.~\cite{Oksendal2007} for further details.

The first diffusion theory was developed by Smoluchowski \cite{Smoluchowski1906} and, independently, Einstein \cite{Einstein1956} at the beginning of the 20th century.  The fundamental object of this theory was the transition probability density, $p_{s,t}(z_{\mathrm{initial}},z_{\mathrm{final}})$ from position $z_{\mathrm{initial}}$ to position $z_{\mathrm{final}}$ between times $s$ and $t>s$.  The individual particle trajectories played no major role in this description of the diffusion process.  In the 1930's Kolmogorov showed that defining a probability measure on the path space is equivalent to specifying all finite-dimensional distributions of a stochastic process, i.e. all joint distributions of the random variables $z(t_1), z(t_2), \dots, z(t_k)$, where $t_1 < t_2 < \dots < t_k$ are some time instants (Kolmogorov extension theorem, for a thorough mathematical treatment see, e.g., Ref. \cite{Durrett2010}).  For the Markov processes, such as processes describing motion of Brownian particles, finite-dimensional distributions are in turn determined by the initial distribution and by the transition probability density mentioned above. This shows, at an abstract level, equivalence of Kolmogorov's general approach and Smoluchowski-Einstein transition probability description.

While the Kolmogorov extension theorem restores the concept of an individual path to its position of central importance in the description of a diffusion process, it still leaves out another fundamental physical ingredient:  infinitesimal evolution law.  This is furnished by the theory of stochastic differential equations (SDE), which describe time evolution of individual paths.  This theory, developed in the 1940's by It\^o \cite{Oksendal2007} (and to some extent also by Gikhman and by Stratonovitch) presents diffusion as a motion of a particle according to a random dynamical law.  Solutions of SDE are Markov processes and, at least in principle, transition probabilities are determined by the equation and, as we have seen, they determine the statistics of the trajectories.

The above remarks are purported to explain the status of the Fokker-Planck equation \cite{Oksendal2007} and its place in the diffusion theory.  This partial differential equation (in mathematics called ``Kolmogorov forward equation'') has the transition density as its solution.  It is thus an infinitesimal approach to diffusion at the level of the transition densities.  As explained above, knowing the solution of the Fokker-Planck equation gives one full knowledge of the diffusion process.  It is thus of utmost importance for a successful implementation of the Fokker-Planck equation method, given an SDE, to be able to produce the correct form of the Fokker-Planck equation from its coefficients.  Before we address this point, we have to discuss the so-called It\^o-Stratonovitch dilemma, which is at the root of the existence of the spurious drift, and deals with the interpretation of SDE \cite{vanKampen2007}.

\begin{table*}
\caption{\label{tab:alpha_table}Correct use of SDE and Fokker-Planck equations for systems coupled to a heat bath.}
\begin{ruledtabular}
\begin{tabular}{ccc}
\boldmath$\alpha$ & \textbf{SDE}  & \textbf{Fokker-Planck equation} \\
\hline
& & \\
$0$
&
$dz = -\frac{F(z)}{\gamma(z)}\,dt+\textcolor{blue}{\frac{\partial}{\partial z}D(z)}\,dt + \sqrt{2D(z)}\,dW$
&
$\frac{\partial}{\partial t}p=\left[\frac{\partial}{\partial z}\frac{F(z)}{\gamma(z)} \textcolor{blue}{- \frac{\partial^2}{\partial z^2}D(z)} +\frac{\partial}{\partial z}\left(\textcolor{blue}{\frac{\partial}{\partial z} D(z)}\right)\right]p$ \\
& & \\
$1/2$
&
$dz = -\frac{F(z)}{\gamma(z)}\,dt+\textcolor{blue}{\frac{1}{2}\frac{\partial}{\partial z}D(z)}\,dt + \sqrt{2D(z)}\,dW$
&
$\frac{\partial}{\partial t}p=\left[\frac{\partial}{\partial z}\frac{F(z)}{\gamma(z)} \textcolor{blue}{- \frac{1}{2}\frac{\partial^2}{\partial z^2}D(x)} +\frac{\partial}{\partial z}\left(\textcolor{blue}{D(z)^{\frac{1}{2}}\frac{\partial}{\partial z} D(z)^{\frac{1}{2}}}\right)\right]p$ \\
& & \\
$1$
&
$dz = -\frac{F(z)}{\gamma(z)}\,dt + \sqrt{2D(z)}\,dW$
&
$\frac{\partial}{\partial t}p=\left[\frac{\partial}{\partial z}\frac{F(z)}{\gamma(z)} +\frac{\partial}{\partial z}\left(\textcolor{blue}{D(z)\frac{\partial}{\partial z}}\right)\right]p$ \\
& & \\
\end{tabular}
\end{ruledtabular}
\end{table*}

\subsection{Stochastic differential equation approach}

We consider the one-dimensional, time-independent SDE
\begin{equation}\label{eq:sde}
dz = \frac{F(z)}{\gamma(z)}\,dt + \sqrt{2 D(z)}\,dW(t),
\end{equation}
where $W(t)$ denotes a Wiener process, i.e. a stochastic process, whose increments are stationary, independent and normally distributed with $W(t) - W(s)$ having mean zero and variance equal $|t-s|$. Such equation describes, for example, the behavior of a Brownian particle in the presence of a position-dependent diffusion coefficient $D(z)$. While only one-dimensional SDE are discussed here, these considerations can be straightforwardly generalized to the multidimensional case.

Eq.~(\ref{eq:sde}) should be interpreted as the integral equation \cite{Oksendal2007}
\begin{equation}\label{eq:sde}
z(T) = \int_0^T \frac{F(z)}{\gamma(z)} \,dt + \int_0^T \sqrt{2 D(z)} \,dW(t).
\end{equation}
However, due to the irregularity of the Wiener process, as well as the solution $z(t)$ of the SDE, the second integral on the right-hand side has several possible interpretations.  More precisely, it is defined as a limit of integral sums,
$\sum_{n=0}^{N-1}\sigma(z(t_j^*)) ( W(t_{j+1}) - W({t_j}) )$,
where $t_j$ are points dividing the interval $[0,T]$ into $N$ equal subintervals and the intermediate points are defined by
$t_j^* = (1-\alpha)t_j + \alpha t_{j+1}$.
A crucial point is that the value of the limit, i.e. of the stochastic integral, depends on the choice of $\alpha$.
Thus, in every applied problem, in addition to the SDE, we have to know the value of $\alpha$ for the mathematical model of the studied phenomenon to be well-defined.  Common choices are $\alpha = 0$ (the It\^o convention), $\alpha = 0.5$ (the Stratonovitch convention) and $\alpha = 1$ (the isothermal, anti-It\^o or backwards-It\^o convention).

The different choices of $\alpha$ are connected to each other by a precise mathematical relationship.  Namely, the above SDE with a given choice of $\alpha$ is equivalent to the It\^o equation ($\alpha = 0$)
\begin{equation}\label{Eq:ModifiedIto}
dz = \frac{F(z)}{\gamma(z)}\,dt + \alpha \frac{d D(z)}{dz}\,dt + \sqrt{2 D(z)}\,dW|_{\alpha = 0}.
\end{equation}
That is, an equation with any choice of $\alpha$ can be rewritten equivalently as an It\^o equation, at the cost of adding an additional term $\alpha \frac{d D(z)}{dz}\,dt$.  This term has been called ``spurious drift'' -- a confusing name since it may suggest its nonphysical character, while,  as we have seen, this additional drift term may be fully observable in a real physical situation \cite{Volpe2010PRL}.

Here, in a nutshell, we see the central problem addressed in this paper:  the choice of $\alpha$ leads in the It\^o form of the equation to an extra drift term, which vanishes only when the diffusion coefficient is constant.  When $D$ is position-dependent, at most one of these values of $\alpha$ can correctly describe the system; such parameter may depend on the system under investigation. However, if the system is strongly coupled to a heat bath, as is the case for a Brownian particle, the spurious drift, and the associated spurious force, are maximal, i.e. $\alpha = 1$ in Eq.~(\ref{Eq:ModifiedIto}). Of course, as shown in Tab.~\ref{tab:alpha_table}, other conventions may also be used adjusting the weight of the spurious drift term; remarkably, for the anti-It\^o or isothermal convention the correction term vanishes \cite{Lau2007}.

\subsection{Fokker-Planck approach}

Since the Fokker-Planck equation is deterministic, its solution involves no randomness and is thus uniquely determined. However, given a SDE, the prescription for writing the associated Fokker-Planck equation depends on the adopted convention, as is shown in Tab.~\ref{tab:alpha_table}.


\end{document}